# Tuning and Optimizing the Finite Element Analysis with Elements of Large Nodal DOF on a Linux Cluster

Ji Wang[*], Lihong Wang[*], Qiang Sun[*], Rongxing Wu[†], Bin Huang[*], Jianke Du[*], and Wei Xiang[#]

[*]Department of Mechanics & Engineering Science, School of Mechanical Engineering and Mechanics, Ningbo University, 818 Fenghua Road, Ningbo, Zhejiang 315211, CHINA

[†]Department of Architectural and Construction Engineering, Ningbo Polytechnic College, 1069 Xinda Road, Beilun District, Ningbo, Zhejiang 315800, CHINA

[#]Department of Mechanical Engineering, School of Mechanical Engineering and Mechanics, Ningbo University, 818 Fenghua Road, Ningbo, Zhejiang 315211, CHINA

**ABSTRACT**

The finite element analysis of high frequency vibrations of quartz crystal plates is a necessary process required in the design of quartz crystal resonators of precision types for applications in filters and sensors. The anisotropic materials and extremely high frequency in radiofrequency range of resonators determine that vibration frequency spectra are complicated with strong couplings of large number of different vibration modes representing deformations which do not appear in usual structural problems.  For instance, the higher-order thickness-shear vibrations usually representing the sharp deformation of thin plates in the thickness direction, expecting the analysis is to be done with refined meshing schemes along the relatively small thickness and consequently the large plane area.  To be able to represent the precise vibration mode shapes, a very large number of elements are needed in the finite element analysis with either the three-dimensional theory or the higher-order plate theory, although considerable reduction of numbers of degree-of-freedom (DOF) are expected for the two-dimensional analysis without scarifying the accuracy. We have successfully implemented the Mindlin plate theory for the analysis of quartz crystal resonators with the finite element method based on both linear and nonlinear formulations for vibration frequencies and mode shapes in the vicinity of the thickness-shear frequency of the fundamental and third-order overtone modes.  The large number of DOF of the linear system resulted from the finite element formulation has challenged the procedure of eigenvalue computation in a specified interval, again which is not encountered often in structural vibrations.  As part of the software development, many different libraries and functions have been utilized in the transformation and eventual evaluation of the eigenvalue problem and the efficiency and accuracy have to be taken into consideration in the solution process.  In this paper, we reviewed the software architecture for the analysis and demonstrated the evaluation and tuning of parameters for the improvement of the analysis with problems of elements with a large number of DOF in each node, or a problem with unusually large bandwidth of the banded stiffness and mass matrices in comparison with conventional finite element formulation. Such a problem can be used as an example for the optimization and tuning of problems from multi-physics analysis which are increasingly important in applications with excessive large number of DOF and bandwidth in engineering.

## 1. Introduction

In the design of quartz crystal resonators, the analysis of structural vibrations of quartz crystal resonators with complications are needed in the determination of vital structural parameters which are closely related to resonator performance and circuit parameters. Unlike usual vibrations of structures we encounter in daily life, vibrations of quartz crystal resonator are different from many aspects with major ones as the relatively higher vibration frequency in the thickness-shear mode and strong anisotropic materials. In addition, there is also a coupling of mechanical deformation and electrical field which further complicates the analysis. As a result, the accurate analysis of vibrations of quartz crystal resonators and calculation of circuit parameters are presented as challenges to the theory of three-dimensional (3D) piezoelectricity and the availability of computing resources. As in similar problems concerning wave propagation in complex materials and structures, essential techniques for the finite element method (FEM) for large scale problems like parallel processing have been utilized. Particularly, one more advantage to take in the case of quartz crystal resonator analysis is the possible adoption of the two-dimensional (2D) Mindlin plate theory which can reduce the dimension of problems within the framework of three-dimensional piezoelectricity. Undoubtedly, it is one of the drastic approaches in certain problems which can be legitimately reduced from the size of the linear problems with the FEM implementation. Consequently, the analysis of quartz crystal resonators has been done with all possible techniques for aggressive reduction and speed-up with modern techniques of FEM and numerical procedures. We shall review the 2D Mindlin plate equations first for a better understanding of the core reduction through the simplification of the 3D problem to a 2D one with the consideration of the geometry of the quartz crystal resonator and its particular vibration mode. The effectiveness of such an approach is based on above features. Then we listed the numerical algorithms utilized in the computational part with emphasis on the eigenvalue extraction which involves the Lanczos algorithm and latest combination of software components for efficiency. The description of the approach we are taking about the ongoing project is to be improved as latest software components are being upgraded.

## 2. FEM Formulation

The generalized Mindlin plate theory is the reduced 3D theory of piezoelectricity for plates vibrating in the thickness-shear mode [1-7]. Such a theory is not familiar to many other physical problems, although the reduction of dimension is one of the major methods in dealing with physical problems of large domains in numerical computation. In essence, the Mindlin plate theory is for the analysis of high frequency vibrations of plates, because more vibration modes can be considered while the number of equations can be selected based on the targeted frequency and mode. Of course, the original objective of the development of Mindlin plate equations is specifically for the design and analysis of quartz crystal resonators. It has made great contribution for the in-depth understanding of the functional mechanism of quartz crystal resonators and its continuous optimization and innovation.

The analysis of vibrations of piezoelectric solids starts from the 3D piezoelectricity theory with equations [2, 8]

$$T_{ij,i} + f_i = \rho \ddot{u}_i, \quad i = 1,2,3, \qquad (1)$$
$$D_{i,i} = q,$$

where $T_{ij}, f_i, \rho, u_i, D_i,$ and $q$ are strains, body forces, density, displacements, electric displacements, and electrical charge, respectively.

The Mindlin plate equations are derived from the generalized expansion of deformation or displacements in power series of the thickness coordinate as [3-9]

$$u_j(x_1, x_2, x_3, t) = \sum_{n=0}^{N} u_j^{(n)}(x_1, x_3, t) x_2^n, j = 1, 2, 3, 4; N - \text{integer}. \qquad (2)$$

where $u_j^{(n)}, x_i (i = 1,2,3), t$ are the $n$th-order displacements, coordinates and time, respectively. It should be mentioned that one of the displacements, usually $u_4$ is designated as the electrical potential. Then by substituting the displacements expressions in (2) into the 3D equations of piezoelectricity, we shall get a set of 2D equations, or plate equations, of infinite order to describe the vibrations as [3-9]

$$T_{ij,i}^{(n)} - nT_{2j}^{(n-1)} + F_j^{(n)} = \sum_{n=0}^{N} B_{mn} \rho \ddot{u}_j^{(m)}, i,j = 1,2,3,4; n = 0,1,2,\cdots, N, \qquad (3)$$
$$D_{i,i}^{(n)} - \overline{D}_2^{(n)} = B_{n0} q,$$

where the variables with superscripts are 2D variables based on the transform process from 3D to 2D. Details of these variables and the derivation process can be easily found in references.

It is emphasized that the Mindlin equations are intended for the calculation of vibration frequencies, as verified by Yong and co-authors [9]. Late analysis of high frequency vibrations of crystal plates has proven that the deformation, or mode shapes, can also be accurately predicted by the Mindlin plate equations. In the analysis, the equations have to be truncated to allow only finite number of equations to be retained, and the implication of the procedure and accuracy have been studied also [8].

The finite element formulation and implementation of the 2D equations have been presented in earlier studies [5-7]. It is obvious that a standard finite element analysis problem has been obtained with unique features: 1) larger bandwidth in the banded stiffness matrices, 2) mass matrix is no longer diagonal, and 3) the calculations of eigenvalues are to be done in a specified interval. All these challenges can be overcome with currently available technology and hardware resources at the cost of computing speed and efficiency. Through the discretization of (3), we shall have a FEM implementation with nodal vector of generalized displacement variables in the length of 4(order + 1), implying 16 variables for the third-order plate equations which are needed for typical analysis. Usual elements like the 4- and 9-node quadrilateral types are used in the development.

For the analysis of quartz crystal resonators, FEM has been tried with the Mindlin plate theory before to solve the above-mentioned challenges with specifically developed tools and software components [3, 5-7, 9]. For instance, in the eigenvalue extraction process the Lanczos method has been utilized and the computing cost issue has been offset by adopting sophisticated techniques for parallelization and preconditioning [10]. Since the quartz crystal resonators are tiny devices fabricated for important but less expensive applications, it is rare to have large investment on the analysis with hardware and developers to improve the computational process. The best approach for the development of computational tool is to use off-the-shelf components and standard cluster computers. This has been major trend of the development from earlier efforts of Motorola supported by the Department of Energy supercomputing initiatives [11-12].

### 3. Software Architecture

With the completion of the finite element formulation, our objective will be on the computation of eigenvalues and eigenvectors of the linear system with aforementioned features. It has been known that since the eigenanalysis has to be done in a given interval, Lanczos algorithm has to be used for the computation of eigenvalues [13, 14]. It was a challenge before and it is still time-consuming today due to the size of the problem is in the million DOF range. Our focus will always be on the speed-up and efficiency of the computing process by utilizing the latest software components for the parallel computation of eigenvalues [10]. In addition, we have to be focused on the verification of the results through some products. More results for indirect comparison from products are presented for the thermal effect, or the frequency-temperature properties of quartz crystal resonators [7]. Latest development in our group has been on the consideration of material viscosity and its effect on the vibrations, which inevitably will also have effects on the circuit parameters of resonators [15]. The challenge is from the fact that the material viscosity will results in a linear system with complex matrices, thus doubling the cost of computation, if the computation of complex eigenvalue and vectors are allowed by all the software components we use. In the future development, we shall keep this requirement and objective in mind for the software updates and choices.

As a typical eigenvalue problem of structural vibrations, the software architecture is shown in Fig. 1. The task is now on the integration of these software components for the operating system and particular hardware platform. In our case, the standard Linux operating system on an Intel-cluster has been provided for the software development and computing. The configuration of computing environment and system alignment are done in accordance with the software requirements.

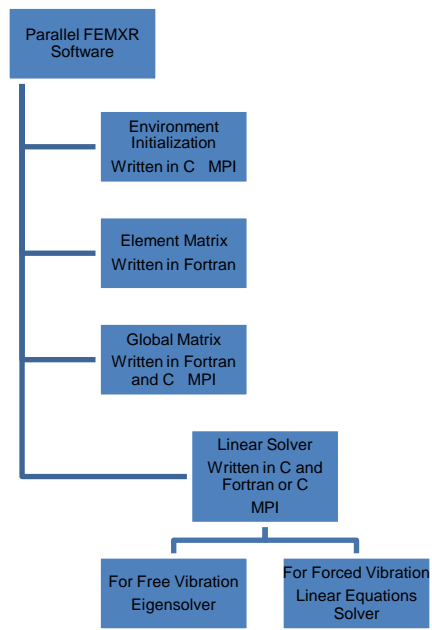

Fig. 1 The architecture of the FEMXR

One particular issue now is the tuning and optimizing of the software because of the components are from different sources and the performance has to be improved with careful configuration of parameters of the platform, operating system, and problem itself. It is a challenge to make the software performance in optimal state for the given resources. In addition, the upgrade of the software could also be a hard task because all the components, as listed in Fig. 2, have to be compatible.

## 4. Optimization and Tuning

Now the tuning and optimizing of the software is important, because we have to make the software usable with different combinations of operating system, CPU maker, root libraries, and interconnections. It is even more important to make the software working on different platforms and environments with acceptable performance. For this reason, we have been examining the initial settings of many system parameters like the storage size, array size, and working space along with the operating system and hardware parameters. Benchmarking tools for system tuning and optimization are also used to aid the process. Given the fact that the focus of our development is on the implementation and results, the optimization and tuning are always on the bottom of the priority list of our development plan. In addition, we lack the knowledge and strategies in modifying some of the software components in accordance with our hardware and operating system for optimal performance. As a result, the effort and scope of tuning and optimization are limited at this stage. We are in the process to gather more experiences on running the software in different environments to make it usable by engineers first. We hope the major effort on tuning and optimization can be made when the software can be stabilized for full functions and suggestions and input from users can be collected and reviewed.

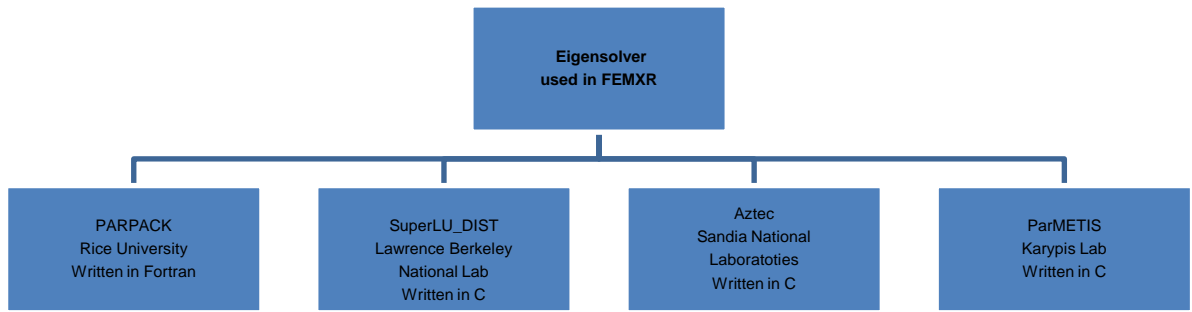

Fig. 2 Major numerical functions of FEMXR

## 5. Conclusions

With continuous efforts on the development and modification, our finite element analysis for the high frequency vibrations of quartz crystal plates and resonators can now be done with software FEMXR developed by ourselves. By performing the analysis with a 2D theory, the numerical problem is significantly reduced in its size with increase on the width of the band of stiffness and mass matrices for a small modification of the characteristics of the eigenvalue problem. Theoretically, the eventual results of the eigenvalues and eigenvectors are very close to 3D analysis based on general-purpose finite element software. The basic functions of the FEMXR include the frequency and mode shapes which can be used as important references in the design of resonators. Further feature of the software also include the coupled analysis of the electrical effect on the piezoelectric material and the distribution of electrical field. Eventually, some electrical parameters of the resonator can also be calculated, which is important for the estimation of resonator properties. Thermal effect of resonators has also been considered. Future improvements including the material and structural viscosity so the complete properties of a resonator and application circuit can be made. We are also interested in improve the visualization capability of the software so the results can be better presented and evaluated. Many of the developments are being discussed with users in the quartz crystal industry.

## Acknowledgments

This research has been supported by government agencies of Zhejiang Province and City of Ningbo under industrial promotion programs and the National Natural Science Foundation of China (Grant No. 11372145). We also thank our corporate sponsors, particularly the TXC Corporation and Hong Kong Crystal Limited, for their continuous financial support and technical collaboration.